\documentclass[two column]{article}

\title{Comment on \lq\lq Charge expulsion and electric field 
       in superconductors"}

\author{Tomio Koyama    \\
Institute for Materials Research, \\
Tohoku University, Sendai 980-8577, Japan}

\begin{document}
\maketitle
\begin{abstract}
We propose a generalized London theory that can correctly describe the 
longitudinal and transverse responses of conventional superconductors to  
an electromagnetic field. The continuity equation is satisfied by use 
of a special gauge for the scalar and vector potentials. Our phenomenological 
theory provides a simple example of the Anderson-Higgs mechanism. 
\end{abstract}


\bigskip\bigskip

In ref.\cite{Hirsch1} Hirsch claimed that hole superconductors should carry 
a surface charge in the ground state even without an applied electric field 
and calculated phenomenologically the inhomogeneous charge appearing inside a 
superconductor. In this comment we point out that the phenomenological 
calculation given in ref.\cite{Hirsch1} for the charge distribution is not 
correct, which includes a misunderstanding of the electrodynamics of a 
superconductor. Since a similar misunderstanding is also seen in recent 
articles \cite{Jan, Hirsch2}, it will be worth presenting the correct 
phenomenological theory based on the generalized London theory for the 
electrodynamics of a superconductor. 

In ref.\cite{Hirsch1} the charge density $\rho(\vec r)$ in a superconductor is 
assumed to satisfy the equation,  
\begin{equation}
\rho(\vec r)= -\frac{1}{4\pi\lambda_L^2}A_0(\vec r), 
\end{equation}
under the Lorenz gauge $\vec\nabla\cdot\vec A(\vec r)+\frac{1}{c}\partial_tA_0
(\vec r)=0$ with $(A_0,\vec A)$ being the scalar and vector potentials. 
In eq.(1) $\lambda_L$ is the \textit{London penetration depth}, which 
indicates that \textit{the screening length of a longitudinal 
electric field in the superconducting state is equal to the London 
penetration depth, not the Thomas-Fermi screening length} $\lambda_{TF}$. 
This point is stressed in ref.\cite{Hirsch1, Jan, Hirsch2} as the remarkable 
electric properties 
in the superconducting state. However, eq.(1) cannot be accepted, because 
it is against the general understanding that the charge screening length is 
still given by $\lambda_{TF}$ in the superconducting state. The charge density 
$\rho(\vec r,t)$ induced by an external electric field in the superconducting 
state can be calculated microscopically within the linear response theory, 
using the formula, $\rho(\vec r,t)=e^2\int dt^\prime\int d\vec r^\prime
K(\vec r-\vec r^\prime,t-t^\prime)A_0(\vec r^\prime,t^\prime)$, where 
$K(\vec r-\vec r^\prime,t-t^\prime)$ is the density-density correlation 
function, i.e., 
$K(\vec r-\vec r^\prime,t-t^\prime)=-i\theta(t-t^\prime)<[\hat n(\vec r,t),
\hat n(\vec r^\prime,t^\prime)]>$. 
In the lowest order (single-loop) approximation for 
$K(\vec r-\vec r^\prime,t-t^\prime)$ of a $s$-wave BCS superconductor one 
obtains the equation in the static limit, 
\begin{equation}
\rho(\vec r)=e^2\sum_{\vec q}K(\vec q)A_0(\vec q){\rm e}^{i\vec q
\cdot\vec r}, 
\end{equation}
where
\begin{equation}
K(\vec q)=-\sum_{\vec k}\frac{E_kE_{k+q}-\epsilon_k
\epsilon_{k+q}+\Delta^2}{E_kE_{k+q}(E_k+E_{k+q})}, 
\end{equation}
with $E_k=\sqrt{\epsilon_k^2+\Delta^2}$. From eq.(3) it follows 
\begin{equation}
K(0)=-\sum_{\vec k}\frac{\Delta^2}{E_k^3}=-2N(0), 
\end{equation}
in the limit $\vec q \rightarrow 0$, where $N(0)$ is the density of states at 
the Fermi level. Note that the gap function $\Delta$ disappears after the 
$k$-integration. Then, one finds the equation valid in the long wavelength 
region,  
\begin{equation}
  \rho(\vec r)= -2e^2N(0)A_0(\vec r)
              = -\frac{1}{4\pi\lambda_{TF}^2}A_0(\vec r). 
\end{equation}
Note that the charge screening length coincides with the Thomas-Fermi 
length $\lambda_{TF}^2=1/(8\pi e^2N(0))$. From this observation one 
understands that the phenomenological theory should be constructed in terms 
of eq.(5) instead of eq.(1), together with the London equation. 
In this case the problem is that what gauge should be chosen, since the 
continuity equation is not fulfilled if one uses the Lorenz gauge, which 
indicates that the Lorenz gauge should be abandoned. Note that the 
substitution of eq.(5) and the London equation into the continuity equation, 
$\partial_t\rho+\vec\nabla\cdot\vec j=0$, leads to the equation as follows, 
\begin{equation}
\vec\nabla\cdot\vec A(\vec r,t) + \frac{\lambda_L^2}{\lambda_{TF}^2}
\frac{1}{c}\partial_tA_0(\vec r,t)=0. 
\end{equation}
We propose that eq.(6) should be regarded as the gauge condition for 
$(A_0,\vec A)$. Notice that the ratio of the screening lengths in eq.(6) 
can be rewritten as 
$\lambda_L^2/\lambda_{TF}^2=c^2/\frac{1}{3}v_F^2\equiv c^2/v_{GB}^2$, 
with $v_F$ being the Fermi velocity, if one assumes $\lambda_{TF}$ and 
$\lambda_L$ in the free electron model. Note that $v_{GB}\equiv v_F/\sqrt{3}$ 
coincides with the velocity of the Goldstone boson in the superconducting 
state. The gauge function $\varphi(\vec r,t)$, which may be considered as 
the phase of the order parameter, i.e., the Goldstone boson, can be 
introduced by the gauge transformation, 
$\vec A\rightarrow \vec A-\frac{\hbar c}{e^\ast}\nabla\varphi$ and 
$A_0\rightarrow A_0 + \frac{\hbar}{e^\ast}\partial_t\varphi$ 
with $e^\ast=2e$ so that the new gauge field also satisfies eq.(6).
It is also noted that the Maxwell-London equations in our theory can be 
derived from the gauge-invariant Lagrangian, 
$$
\mathcal{L}=\frac{1}{8\pi e^{\ast 2}\lambda_L^2}\Bigl\{
\frac{c^2}{v_{GB}^2}\bigl(e^\ast A_0+\hbar\partial_t\varphi\bigl)^2
$$
\begin{equation}
-\bigl(e^\ast \vec A - \hbar c\vec\nabla\varphi\bigl)^2\Bigl\}
+\frac{\vec E^2}{8\pi}-\frac{\vec B^2}{8\pi}. 
\end{equation}
This Lagrangian describes the massive plasma modes. Under the gauge condition 
(6) one can obtain the dispersion relations, 
$\omega_T(p^2)=\omega_p^2(1+\lambda_L^2p^2)$ for the transverse one and 
$\omega_L(p^2)=\omega_p^2(1+(v_{GB}^2/c^2)\lambda_L^2p^2)$ 
for the longitudinal one with $\omega_p=c/\lambda_L$. 
In the neutral case i.e., $e^\ast=0$, the Lagrangian is reduced to the one 
for the massless scalar field, 
$\mathcal{L}=\hbar c^2/(8\pi e^{\ast 2}\lambda_L^2)\Bigl[
v_{GB}^{-2}\bigl(\partial_t\varphi\bigl)^2-\bigl(\vec\nabla\varphi\bigl)^2
\Bigl]$. Then, eq.(7) indicates that the Goldstone boson $\varphi$ is absorbed 
into the gauge field $(A_0,\vec A)$, i.e., the \textit{Anderson-Higgs 
mechanism}\cite{Wein}. Thus, our phenomenological theory provides the 
consistent description for the electrodynamics of a superconductor, in which 
the charge screening length is given by the Thomas-Fermi length. 
The gauge condition given in eq.(6) was proposed in ref.\cite{Matumoto} 
and is called the \textit{phason gauge}.


\end{document}